# Bandwidth selection with a frequency-domain version of the AIC

Erhard Reschenhofer[1]

### Abstract

When it comes to estimating an unknown spectral density as simply and reliably as possible, parametric spectral density estimation using AR models and order selection via AIC is the method of choice. In contrast, no standard method has yet emerged for automatic nonparametric spectral density estimation, and there seems to be little willingness to weigh the advantages and disadvantages of different risk functions and the various methods for estimating them on a case-by-case basis, particularly because it is unclear whether the effort is even worthwhile without concrete prior information about the unknown spectral density. As a result, subjective visual methods are still widely used in practice to determine the appropriate smoothing parameter for a nonparametric estimation. This article aims to encourage the increased use of objective automatic methods by presenting evidence that using what is arguably the simplest and most straightforward frequency-domain version of the AIC for the automatic determination of an appropriate bandwidth enables results that are comparable to those obtained using the standard parametric approach. This evidence is based on both real-world time series and synthetic time series with spectral densities of varying complexity.

**Keywords:** nonparametric spectral density estimation, frequency-domain AIC, automatic periodogram smoothing, unbiased estimation, predictive log likelihood, weighted moving averages.

## 1. Introduction

The use of automatic model selection methods is common in time series analysis, not only in the time domain but also in the frequency domain, such as in the parametric estimation of spectral densities. Indeed, the AR spectral density obtained by minimizing the model selection criterion AIC usually provides a reasonable smoothing of the periodogram. However, if we want to get a second opinion at the click of a button, we cannot just replace the AR models with the more versatile ARMA models, because the latter require a certain degree of supervision. Firstly, when estimating higher-dimensional ARMA models with standard statistical software, one quickly reaches numerical limits when the optimization algorithm used either fails to converge at all or gets stuck at some local optimum. Secondly, estimated ARMA spectral densities sometimes exhibit completely "unnatural" spikes that not only violate the principle "natura non facit saltus" but also result in very poor goodness of fit when measured uniformly rather than pointwise. These spikes occur when the ARMA model attempts to account for a rather abrupt rise/fall in the level of the periodogram at a specific frequency by including a pair of complex-conjugate roots with approximately that frequency and amplitude close to 1 in the numerator/denominator polynomial and another pair with a slightly higher frequency and amplitude also close to 1 in the denominator/numerator polynomial.

[1] *Retired from University of Vienna, E-mail: erhard.reschenhofer@univie.ac.at*

Nonparametric spectral density estimators would be an obvious alternative, yet, for some reason, they are rarely used in fully automated settings. In the simplest case, where the spectral density at any given frequency is estimated by an average of neighboring periodogram ordinates (rectangular window), the number of ordinates used - or the difference between the lowest and highest frequencies used (bandwidth) - determines the smoothness of the estimator. In the more general case of weighted averages, the bandwidth is defined as the bandwidth of a rectangular window with the same (equivalent) degrees of freedom. As always, there is a trade-off between bias and variance here as well. The bias will increase and the variance will decrease as the bandwidth increases. In practice, to find the right balance between bias and variance, one relies on a visual comparison of estimates obtained using different bandwidths. The focus there is on the stability of the estimated spectral density as the window is gradually closed ("window closing").

The fact that automatic methods for bandwidth selection are not used more frequently is partly due to the fact that none of these methods has established itself as a standard - in the same way that AIC has for parametric spectral analysis. Unfortunately, it is not enough to simply choose a particular type of bandwidth selection method. Even if we decide, for example, to determine the bandwidth by minimizing a cross-validatory estimate of the mean integrated squared error (MISE), we still need to choose a specific definition of the MISE (Hurvich, 1985, considered three different definitions) as well as a specific cross-validation method (Robinson, 1991, established consistency for a leave-two-out modification of the leave-one-out cross-validation method proposed by Hurvich, 1985, Beltrao and Bloomfield, 1987). Finally, there is also the option to try smoothing the log periodogram instead of the periodogram (Wahba, 1980; Lee, 1997). It almost seems as though we would need some additional criterion just to choose the bandwidth selection method itself. In order to avoid getting involved in something like this, we count on AIC working in the frequency domain as well, and will therefore exclusively use the frequency-domain counterpart of AIC for bandwidth selection.

The success of AIC is not based so much on it having specific nice theoretical properties (see, e.g., Shibata, 1980, 1981) - despite certain flaws (Kabaila, 2002) - but rather on extensive positive experience in practical application. The purpose of this article is therefore not to show that a particular bandwidth selection method is optimal in a certain sense under certain assumptions in certain situations, but simply to demonstrate that the "natural" frequency-domain version of AIC is just as easy to use and automatically provides nonparametric estimates that are comparable to the parametric ones. To this end, both real and synthetic time series are used. The frequency-domain AIC is explained in the next section. Sections 3 and 4 present the empirical results and the results of the simulations, respectively. All calculations are performed using the statistical software R (R Core Team, 2024). Section 5 concludes.

## 2. Automatic smoothing of the periodogram

When attempting to smooth the periodogram

$$I_k = \frac{1}{2\pi n}\left|\sum_{t=1}^{n} y_t e^{-i\omega_k t}\right|^2 \tag{1}$$

of a time series $y_1,\ldots,y_n$ at the Fourier frequencies $\omega_k = 2\pi k/n$, $k = 1,\ldots,m = [(n-1)/2]$, using (weighted) moving averages, problems naturally arise at the edges - that is, near the frequencies $0$ and $\pi$. These problems can be easily avoided by switching to the representation

$$I_k = \frac{1}{2\pi} \left( \hat{\gamma}_0 + 2 \sum_{j=1}^{n-1} \hat{\gamma}_j cos(\omega_k j) \right) \quad (2)$$

and achieving smoothing by truncating the sum in (2) and shrinking the sample autocovariances

$$\hat{\gamma}_j = \frac{1}{n} \sum_{t=1}^{n-|j|} \left( y_t - \bar{y} \right) \left( y_{t+j} - \bar{y} \right) \quad (3)$$

toward 0. However, since the two smoothing methods are closely related and the first one is more convenient for us in the following, we will stick with that one. Regarding the issue mentioned above, only a minor modification is needed. At the edges, the periodogram can be extended by taking advantage of its periodicity to allow for symmetric averaging there as well. To be more precise, what is actually being exploited here is not the periodicity of the cosine function, but its symmetry at the points 0 and $\pi$.

To simplify the presentation of the theoretical concepts, let us first assume that the stationary process $y$ with spectral density $f$ is Gaussian white noise, because in this case we can make precise statements rather than merely approximate ones - for example, regarding the distribution of the periodogram at the Fourier frequencies, i.e.,

$$J_k = I_k / f(\omega_k) \sim Exp(1),\ 2J_k \sim \chi_2^2,\ 2\left( J_{k-r} + \dots + J_{k+r} \right) \sim \chi_{2(2r+1)}^2 \quad (4)$$

Estimating $f_k = f(\omega_k)$ by the weighted average

$$\hat{f}_k = \sum_{s=-r}^{r} w_s I_{k+s}\ , \quad (5)$$

where

$$w_{-r} + \dots + w_r = 1,\ w_{-s} = w_s \text{ for } s = 0,\dots,r, \quad (6)$$

we obtain

$$E\ \hat{f}_k / f_k = 1,\ Var\ \hat{f}_k / f_k = \sum_{s=-r}^{r} w_s^2 = w_0^2 + 2 \sum_{s=1}^{r} w_s^2, \quad (7)$$

and in particular for the rectangular window

$$\breve{w}_s = 1/(2r+1), \quad (8)$$

the triangular window

$$\check{w}_s = (1 - |s|/r)/r, \quad (9)$$

and the cosine window

$$\tilde{w}_s = (1 + cos\,(s\pi/r)/(2r), \quad (10)$$

$$Var\ \breve{f}_k / f_k = 1/(2r+1), \quad (11)$$

$$Var\ \check{f}_k / f_k = 1/r^2 + (r-1)(2r-1)/3r^3 = \left( 2r^2 + 1 \right) / 3r^3, \quad (12)$$

and

$$Var\, \tilde{f}_k / f_k = 0.75/r, \quad (13)$$

respectively. Apart from a constant factor, the estimator $\breve{f}_k$ has a $\chi^2$ distribution with $\breve{\nu} = 2\,(2r+1)$ degrees of freedom, i.e.,

$$\breve{\nu} = 2 \Big/ \sum_{s=-r}^{r} \breve{w}_s^2. \quad (14)$$

Analogously, we get for the triangular window the “equivalent degrees of freedom”

$$\check{\nu} = 2 \Big/ \sum_{s=-r}^{r} \check{w}_s^2 = 6r^3 \Big/ \left( 2r^2 + 1 \right), \quad (15)$$

and for the cosine window

$$\tilde{\nu} = 2 \Big/ \sum_{s=-r}^{r} \tilde{w}_s^2 = 8r/3. \quad (16)$$

To avoid evaluating the quality of a model (or smoothing procedure) based on data that were already used to estimate the model, we consider the "predictive Whittle-log likelihood"

$$L(I^*|\hat{f}) = -\sum_{k=1}^{m}\left(\log(\hat{f}_k) + I_k^*/\hat{f}_k\right) \tag{17}$$

instead of the standard one, where $I_1^*,\dots,I_m^*$ is an independent sample. In the case of the rectangular window, we have

$$E\left(I_k^* - I_k\right)/\check{f}_k = (2r+1)\,E\left(I_k^* - I_k\right)/(I_{k-r} + \dots + I_{k+r})$$

$$= \tfrac{\check{\nu}}{2}E\left(2J_k^* - 2J_k\right)/(2J_{k-r} + \dots + 2J_{k+r}) \qquad \text{(because } f_1 = \dots = f_m\text{)}$$

$$= \tfrac{\check{\nu}}{2}\left(\tfrac{2}{\check{\nu}-2} - \tfrac{2}{\check{\nu}}\right) = \tfrac{2}{\check{\nu}-2}. \tag{18}$$

The smoothness of the spectral density estimator $\check{f}$ increases with the number of degrees of freedom. When minimizing the negative log likelihood, overfitting (i.e., insufficient smoothing) can be prevented by including a penalty term that decreases as the number of degrees of freedom increases. A suitable degree of smoothness could be determined by minimizing

$$-L\left(I\middle|\check{f}\right) + \tfrac{2m}{\check{\nu}-2} = \sum_{k=1}^{m}\log(\check{f}_k) + \sum_{k=1}^{m} I_k/\check{f}_k + \tfrac{2m}{\check{\nu}-2} \tag{19}$$

or, more generally,

$$C(\hat{f}) = \sum_{k=1}^{m}\log(\hat{f}_k) + \sum_{k=1}^{m} I_k/\hat{f}_k + \tfrac{2m}{\hat{\nu}-2} \tag{20}$$

with respect to $r$, where $\hat{f}_k = \sum_{s=-r}^{r} w_s I_{k+s}$ and $\hat{\nu} = 2/\sum_{s=-r}^{r} w_s^2$.

Since the expected value

$$E\ \log(\check{f}_k) = -\,log(\check{\nu}) + E\ log(\underbrace{2I_{k-r} + \dots + 2I_{k+r})/f_k}_{\sim\chi^2_{\check{\nu}}} + log(f_k)$$

$$= -\,log(\check{\nu}) + log(2) + \psi(\check{\nu}/2) + log(f_k)$$

$$= \psi(\check{\nu}/2) - log(\check{\nu}/2) + const$$

$$\approx -\frac{1}{2\left(\frac{\check{\nu}}{2}\right)} - \frac{1}{12\left(\frac{\check{\nu}}{2}\right)^2} + \frac{1}{120\left(\frac{\check{\nu}}{2}\right)^4} - \frac{1}{252\left(\frac{\check{\nu}}{2}\right)^6} + const$$

$$\approx -\tfrac{1}{\check{\nu}} + const \tag{21}$$

is, apart from an additive constant, negative and approaches 0 as the degrees of freedom increase, the first term on the right-hand side of (19) favors overfitting, but it is still smaller in absolute value than the positive and decreasing penalty term. Here, $\psi$ denotes the digamma function.

Although the constancy of the rectangular window and the constancy of the spectral density played an important role in deriving (19), we will use the general criterion (20) also for other windows and nontrivial spectral densities. Similarly, in the derivation of the most popular model selection criterion, namely AIC, the assumption that the trivial model of dimension 0 is true was used to determine by how much, on average, the negative log likelihood (or the squared prediction error) of the model decreases when the dimension is (unnecessarily!) increased (see Rothman, 1968; Akaike, 1969; Akaike, 1973). Indeed, the AIC is just an asymptotically unbiased estimator of minus 2 times the expected predictive log-likelihood (for unbiasedness in small samples see Sugiura, 1978). When comparing the estimates obtained for different models, one implicitly assumes that they are all correctly specified, in particular also the trivial model of dimension 0 - otherwise, unbiasedness would no longer apply generally. However, it turned out that further developments of AIC, which do not rely on this unrealistic assumption (Sawa, 1978; Reschenhofer, 1999), usually perform similarly to AIC because in the critical cases, where the misspecification is most evident and the penalty terms differ the most, the likelihood term is decisive anyway (see Reschenhofer, 1999).

A far greater problem with AIC-like criteria is that they are only designed for nested models and are not suitable for choosing between arbitrary models. If, in the latter task, the set of candidate models is too large, even the BIC (Schwarz, 1978) loses its most important property, namely its consistency (see, e.g., Reschenhofer, 2015). Fortunately, both the choice of the AR order in the parametric case and the choice of the bandwidth in the nonparametric case are nested problems, because in each case there is a natural ordering.

## 3. Applications

The motivation for this paper stems from a recent study (Reschenhofer, 2026) in which a nonparametric alternative to automatic parametric spectral analysis would have been a useful addition. That study, taking into account exogenous variables such as solar variations and El Niño, found evidence that global warming is accelerating and also that the breach of the 1.5°C Paris Agreement target is imminent (but has not yet occurred). Figures 1.a, 1.c, and 1.d show the global mean surface temperature (GMST) from January 1850 to December 2025, monthly means of the daily total sunspot numbers from January 1749 to February 2026, and the Oceanic Niño Index (ONI) from January 1950 to December 2025, respectively. Figure 1.b shows the detrended GMST from January 1964 to December 2025. Detrending was performed under the null hypothesis of "no acceleration", i.e, no break in the trend, by subtracting a fitted linear trend. Here data prior to 1964 have been omitted to avoid obvious but, in this context, irrelevant breaks. The monthly GMST was obtained from the HadCRUT5 surface temperature dataset (Morice et al., 2021; https://crudata.uea.ac.uk/cru/data/temperature), the mean sunspot numbers from the World Data Center SILSO (WDC-SILSO), Royal Observatory of Belgium (Licence: CC BY-NC: https://creativecommons.org/licenses/by-nc/4.0, https://doi.org/10.24414/qnza-ac80, https://www.sidc.be/SILSO/datafiles), and the ONI from the Climate Prediction Center (CPC; https://www.cpc.ncep.noaa.gov).

In addition to these climate time series, selected economic and financial time series will also be analyzed. The first two are macroeconomic time series, namely the quarterly real US Gross Domestic Product (GDPC1; in billions of chained 2017 Dollars, seasonally adjusted) from 1947.I to 2026.I (Source: U.S. Bureau of Economic Analysis) and the monthly unemployment rate (UNRATE; in percent, seasonally adjusted) from January 1948 to March 2026 (Source: U.S. Bureau of Labor Statistics). Both series were retrieved from FRED, Federal Reserve Bank of St. Louis (https://fred.stlouisfed.org). The missing UNRATE value for October 2025 has been replaced with the subsequent value for November 2025. The third time series is the exchange rate Euro to US Dollar (EURUSD=X) from 2003-12-01 to 2026-05-08, which has been downloaded from Yahoo Finance (https://finance.yahoo.com). This time series is more interesting than other financial time series such as (log) stock prices as it exhibits a "milder form of nonstationarity" and the spectral density of its first differences is not just flat. Figure 2 shows the differenced log GDP, the unemployment rate, the exchange rate and the differenced exchange rate, respectively.

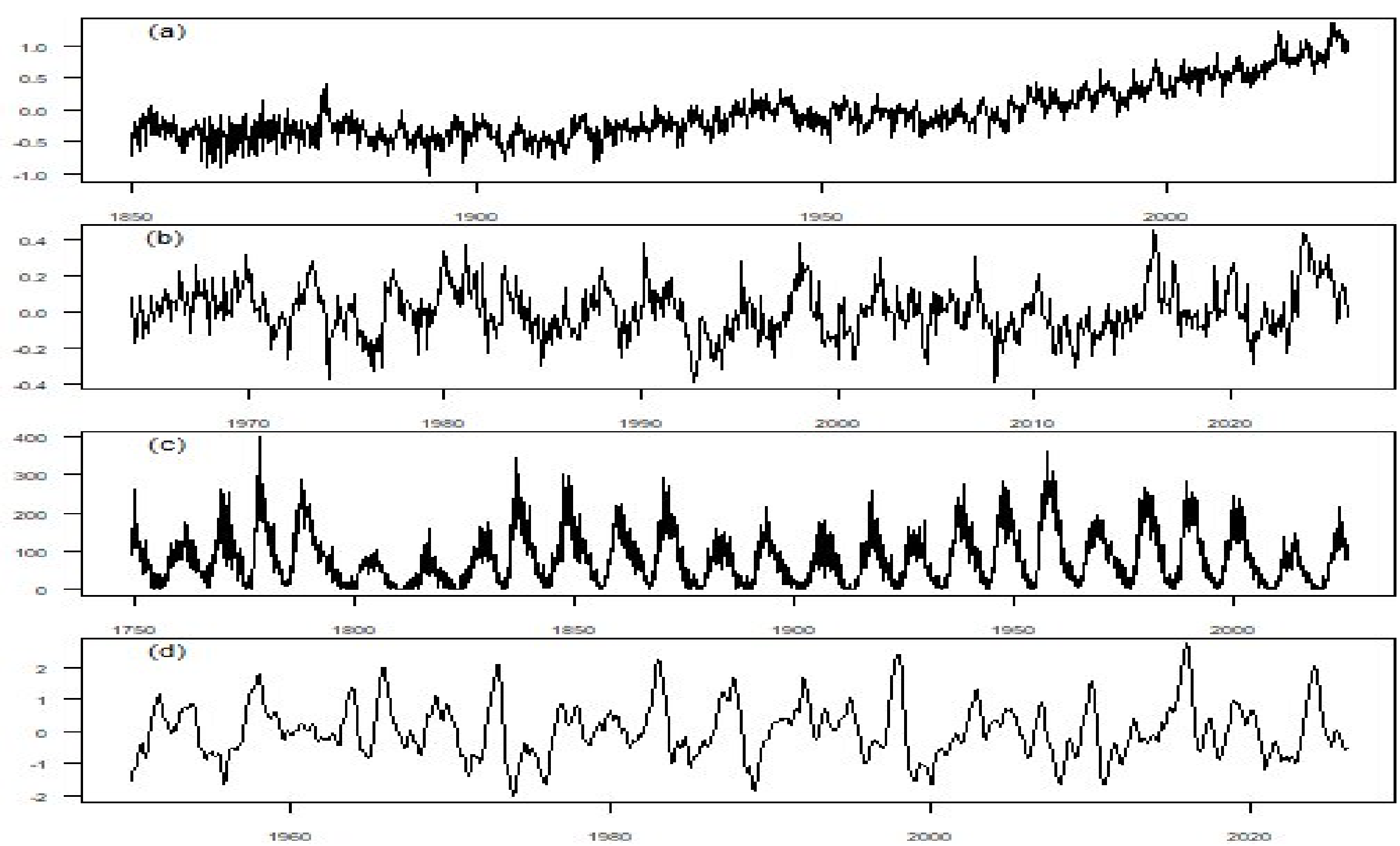

Figure 1: (a) Monthly GMST from January 1850 to December 2025 (HadCRUT5)
(b) Detrended GMST from January 1964 to December 2025
(c) Monthly means of daily total sunspot numbers 1749.01 – 2026.02 (WDC-SILSO)
(d) Monthly ONI from January 1950 to December 2025 (CPC)

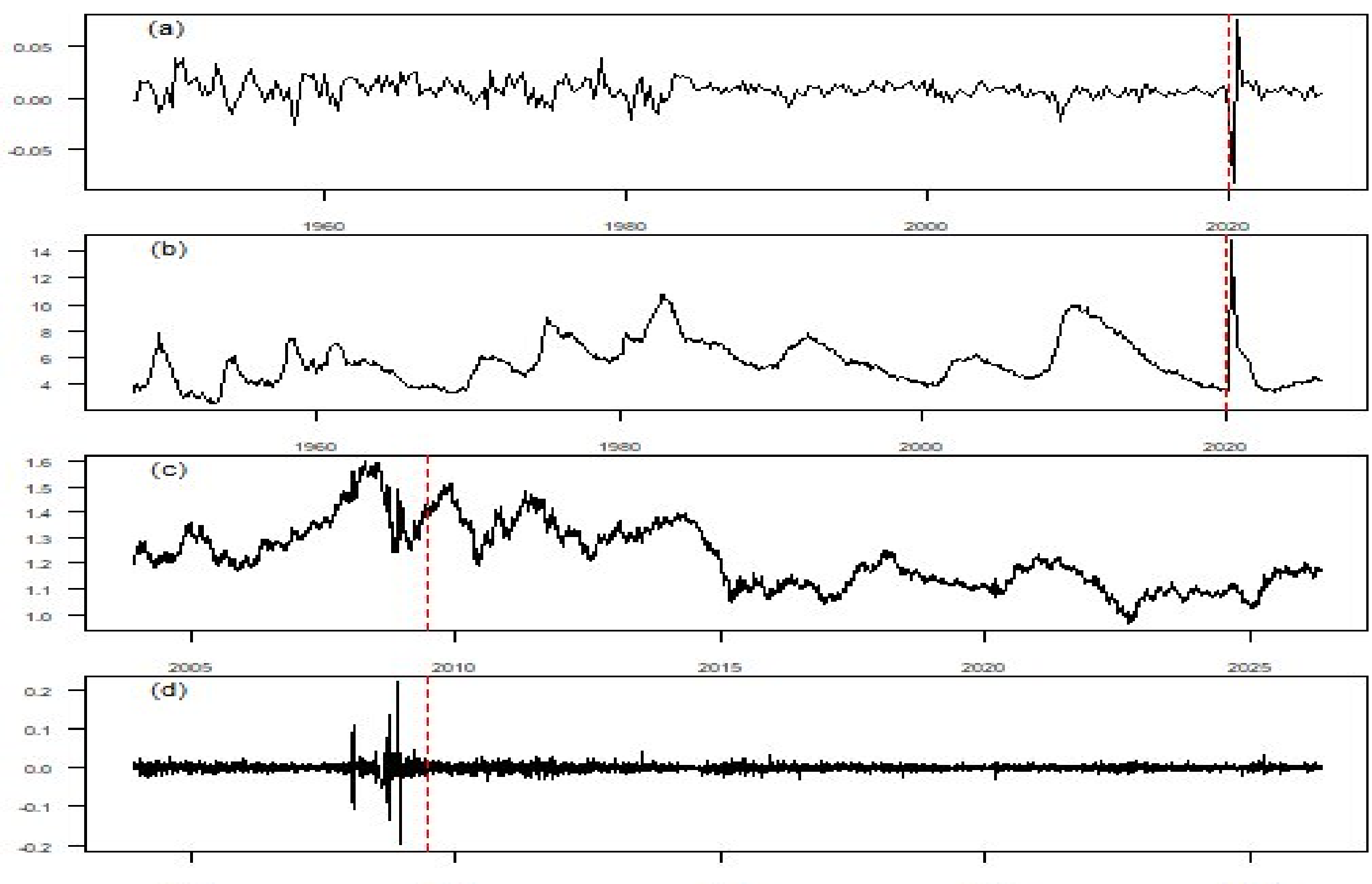

Figure 2: (a) Quarterly GDPC1 from 1947.I to 2026.I (Bureau of Economic Analysis)
(b) Monthly UNRATE from January 1948 to March 2026 (Bureau of Labor Statistics)
(c) Daily EURUSD=X from 2003-12-01 to 2026-05-08 (Yahoo Finance)
(d) Differenced EURUSD=X

The latter four time series contain outliers that need to be addressed. In the case of the macroeconomic time series, extreme outliers occur during the coronavirus pandemic, which began with an outbreak in China in December 2019, and in the case of the financial time series they occur during the global financial crisis, which began in early 2007 with the collapse of the U.S. subprime mortgage market and ended in 2009. Basically, there are two options. Either we address the outliers directly - for example, by scaling them down - or indirectly by using only robust methods to analyze the data. The first option is, by its very nature, somewhat arbitrary, and the second is out of the question for us because the methods have already been specified in the previous section. Fortunately, our focus is not on the data themselves but on the methods. The data are used solely to illustrate differences in performance, so we can simply trim the time series - either by not starting before mid-2009 or by ending before 2020 (see the dashed red lines in the figures).

When applying the various automatic spectral density estimators to the climate time series, it turns out that the nonparametric estimators do not differ significantly from one another or from the AR estimator (see Figure 3). As expected, it essentially comes down to the bandwidth, and the shape of the window does not matter much. Only in the case of the mean sunspot numbers does the simplest window - namely, the rectangular one - stand out noticeably. In this case, as well as in the case of the El Niño series, the ARMA estimator also deviates significantly. However, the ARMA estimates should generally be treated with caution, as convergence problems with the optimization algorithm occur relatively frequently, especially when more than three lags are included.

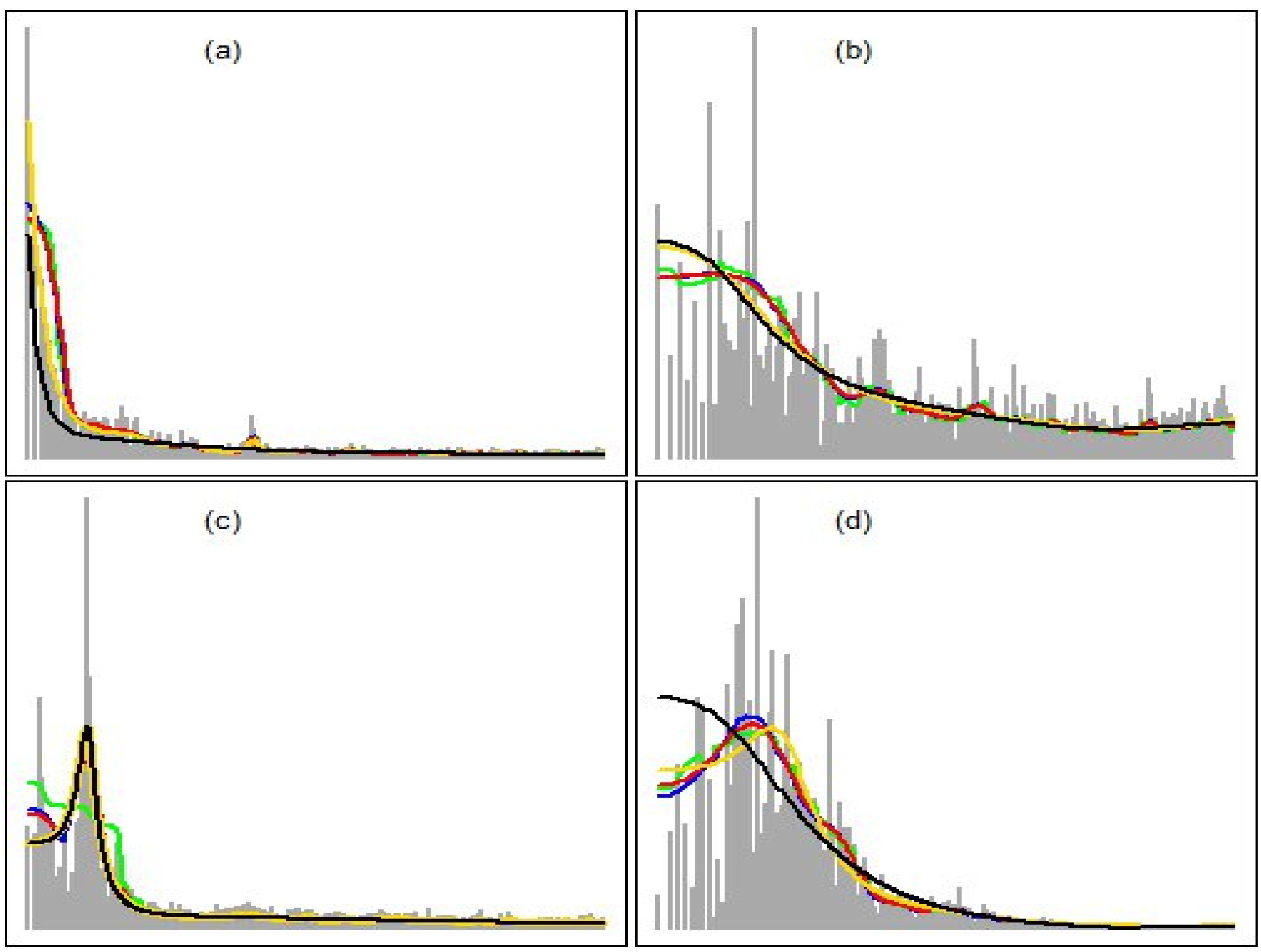

Figure 3: Periodograms (gray) of (a) monthly GMST, (b) detrended GMST, (c) monthly sunspot numbers, and (d) monthly ONI together with automatically estimated spectral densities (ARMA: black, AR: yellow, rectangular: green, triangular: blue, cosine: red). For better visibility, the periodograms and the spectral densities as well as the frequencies were subjected to a square root transformation.

Figures 4 and 5 show the spectral density estimates obtained for the original macroeconomic and financial time series and for the trimmed ones, respectively. Again, there is a high degree of agreement among the competing estimators. The largest discrepancies occur when the ARMA estimator models a sudden drop in the level of the periodogram by a pair of complex-conjugate roots with frequency 0.73 and amplitude close to 1 in the denominator polynomial and another pair with frequency 0.77 and amplitude also close to 1 in the numerator polynomial, both in the case of the differenced log GDP and in the case of the trimmed differenced log GDP. Also noteworthy is the unusually volatile AR estimate in the case of the differenced exchange rate, which actually looks more like what one would imagine a traditional nonparametric estimate to look like. However, that changes immediately as soon as the outliers are removed by trimming the time series. This measure also significantly reduces the volatility of the non-parametric estimates. With the exception of the simple rectangular window, their fluctuations are then much smaller than the recurring spike in the ARMA estimate, which is once again caused by almost canceling roots near the unit circle.

The potential power deficiency in the low-frequency range, which may be discernible in the periodogram of this time series (see Figure 5.d), is ignored by all estimators. Its actual presence could be an indication of overdifferencing. Of course, overdifferencing does not mean that the original time series was already stationary, but merely that stationarity could have been achieved already with fractional differencing of order $d < 1$. Indeed, if we plot $\log(I_k)$ against $-2\log(\sin(\omega_k/2)), k = 1,\ldots,20$ (see Geweke and Porter-Hudak, 1983) for the trimmed exchange rate, we get a slope (see Figure 6) that appears steeper than 0.5 (green reference lines) but flatter than 1 (red reference lines).

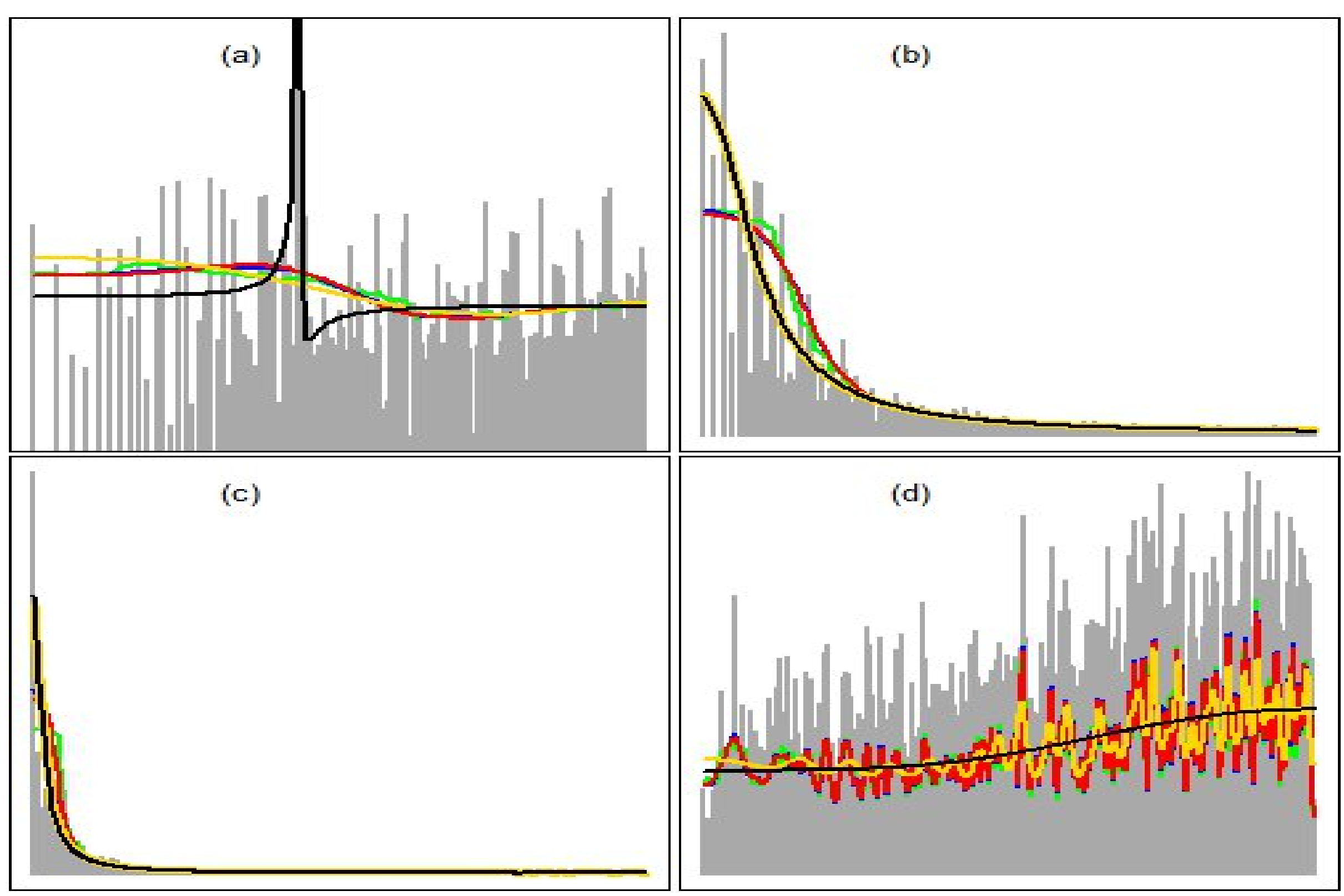

Figure 4: Periodograms (gray) of (a) quarterly GDPC1, (b) monthly UNRATE, (c) daily EURUSD=X, and (d) differenced EURUSD=X together with automatically estimated spectral densities (ARMA: black, AR: yellow, rectangular: green, triangular: blue, cosine: red). For better visibility, the periodograms and the spectral densities as well as the frequencies were subjected to a square root transformation.

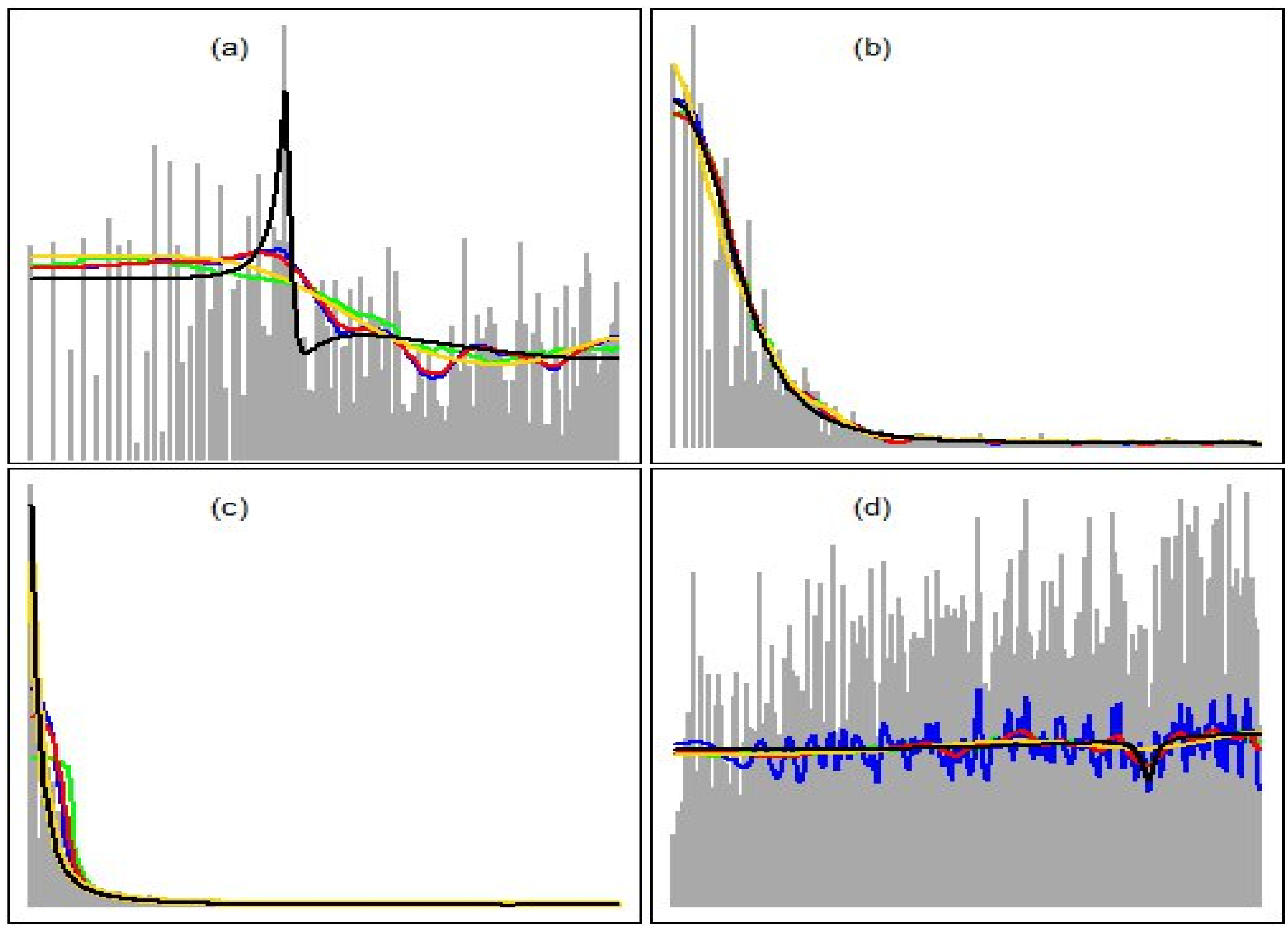

Figure 5: Periodograms (gray) of trimmed (a) quarterly GDPC1, (b) monthly UNRATE, (c) daily EURUSD=X, and (d) differenced EURUSD=X together with automatically estimated spectral densities (ARMA: black, AR: yellow, rectangular: green, triangular: blue, cosine: red). For better visibility, the periodograms and the spectral densities as well as the frequencies were subjected to a square root transformation.

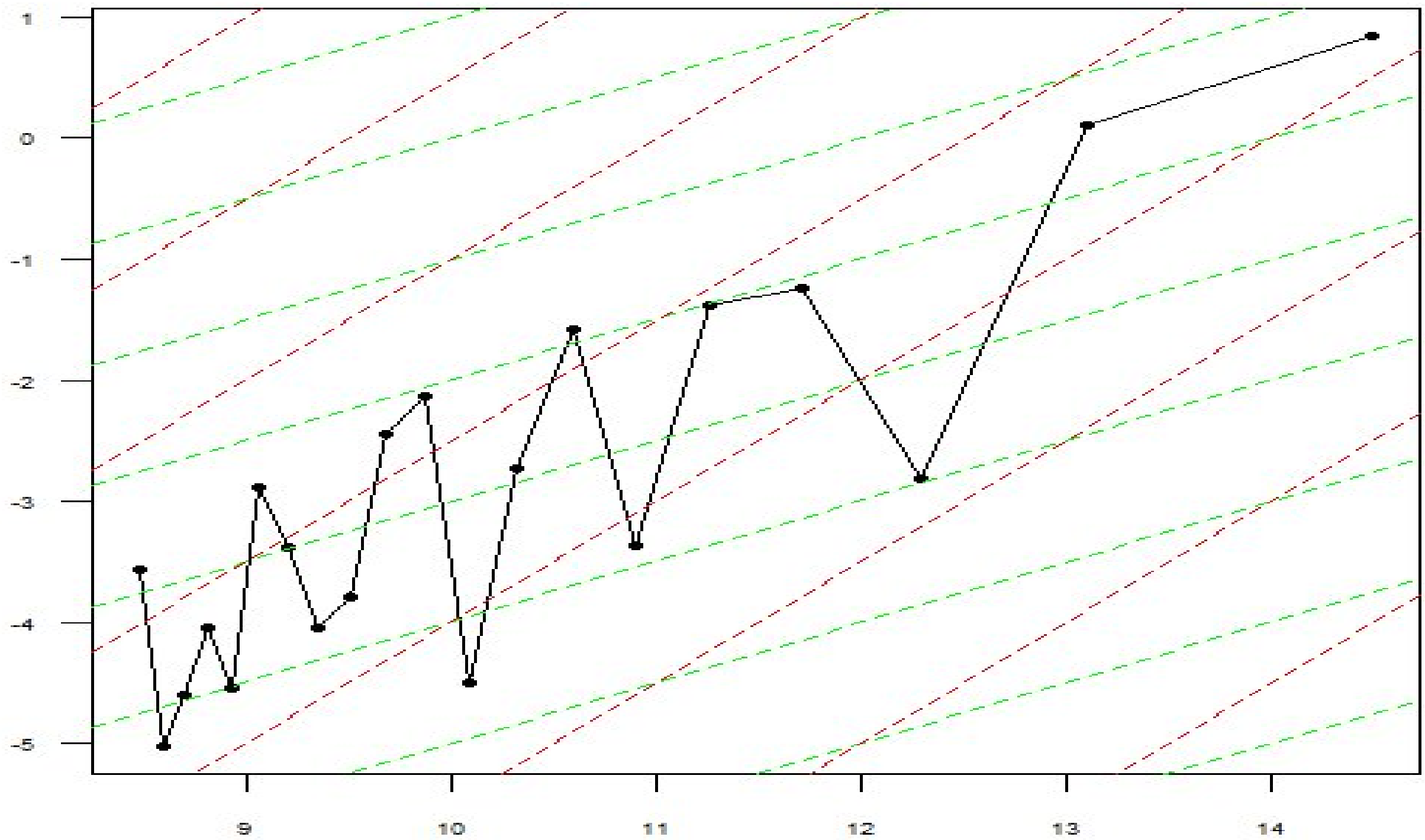

Figure 6: The log periodogram of the trimmed daily EURUSD=X at the first twenty Fourier frequencies is plotted against $-2\log(\sin(\omega_k/2))$, $k = 1,\ldots,20$, together with reference lines with slope 0.5 (green) and 1 (red), respectively.

## 4. Simulations

To generate the synthetic time series, we use MA models instead of AR models so as not to give the AR estimators too much of an advantage. Since real-world time series, particularly those from the field of economics, often have very boring spectral densities, we also include multimodal spectral densities here. The simplest way to achieve this is with an MA model of the form

$$y_t = \theta u_{t-q} + u_t = (1 + \theta L^q)\, u_t = \Theta(L) u_t\,, \tag{22}$$

where the innovations $u_t$ are i.i.d. $N(0,1)$. The number of peaks increases as q increases, e.g., for $q = 6$, the polynomial

$$\Theta(z) = 1 + \theta z^q \tag{23}$$

has three pairs of complex conjugate roots at the frequencies $\pi/6, \pi/2, 5\pi/6$, if $\theta > 0$, and two pairs at $\pi/3$, $2\pi/3$ (as well as two real roots at 0 and $\pi$), if $\theta < 0$.

Figure 7 shows the spectral densities of the data generating MA processes (22) with $\theta = -0.5, 0.5$ and $q = 1,\ldots,6$, together with spectral density estimates obtained from single realizations of length $n = 200, 2000$. In general, there are no major differences between the nonparametric and the parametric estimates. For better visibility, only the estimates obtained with the cosine window are shown here. However, in the simulation study - with 1,000 realizations per MA model and sample size – the rectangular window and the triangular window are included as well (see Table 1).

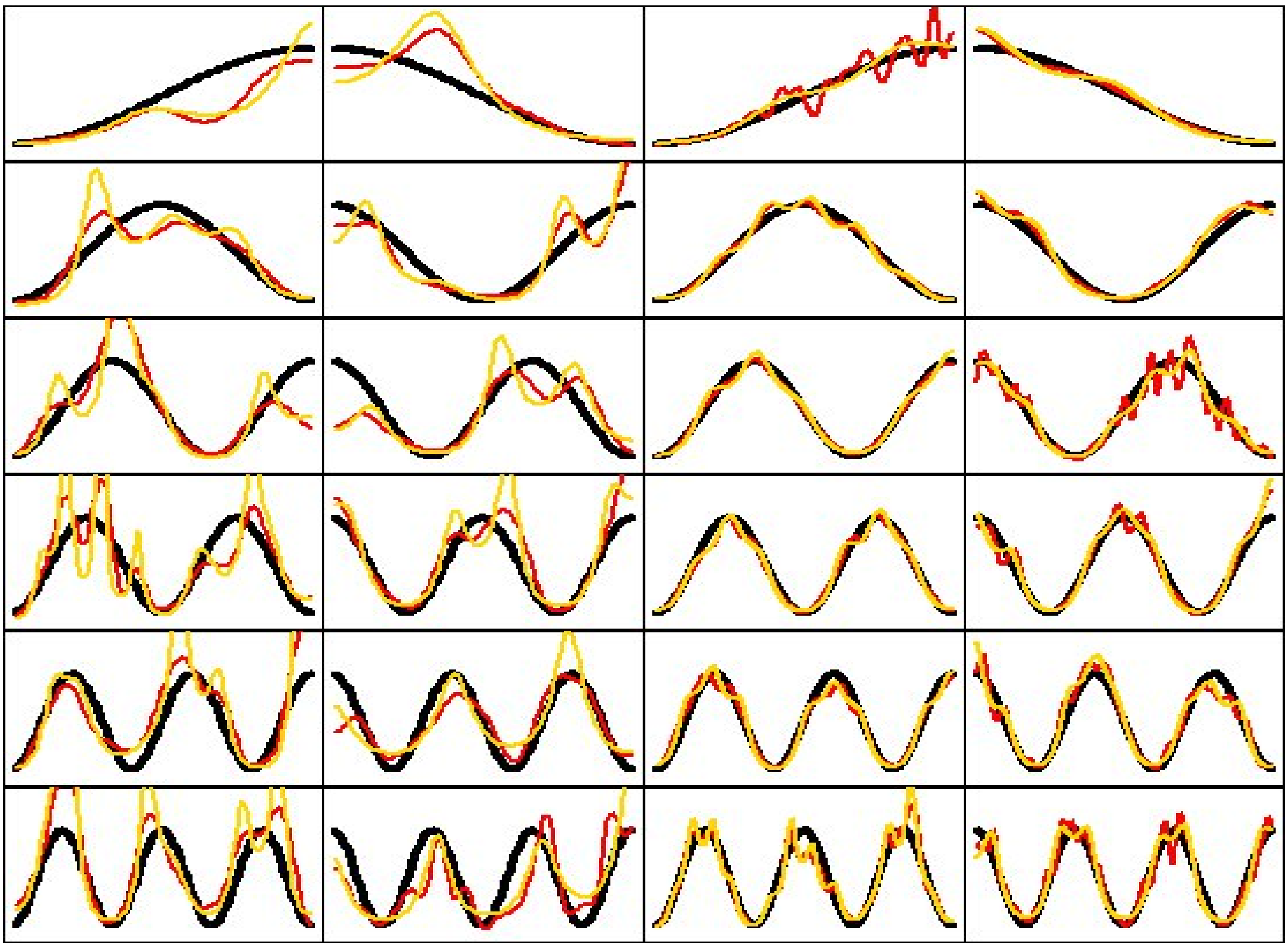

Figure 7: Spectral densities (black) of data generating MA processes (22) with parameter $\theta = -0.5$ (columns 1 and 3) or $\theta = 0.5$ (columns 2 and 4) together with estimates (AR: yellow, cosine: red) obtained from single realizations of length $n = 200$ (first two columns) or $n = 2000$ (last two columns).

For the summary of the simulation results, we need some measure of the quality of the fit. Being aware that the choice of this measure has a major impact on the ranking of the competing estimators, we opt for a reasonably robust measure, namely the sum of absolute residuals or, equivalently, the mean of the absolute residuals

$$GF\left(\hat{f}\middle|f\right) = \frac{1}{m}\sum_{k=1}^{m}\left|\hat{f}_k - f_k\right|. \tag{24}$$

Using this goodness-of-fit measure, we find that the selection criteria work (see Table 1.a). Overall, the data-driven choices perform better than the deterministic ones, such as $p = [\sqrt{m}]$ and $r = [\sqrt{m}]$, although not by as much as might have been expected. Admittedly, a bit of data snooping might have helped. If the average selected AR order in the most complex case ($q = 6$) is just under 10 for $m = 100$ and just over 20 for $m = 1000$ (see Table 1.b), then choosing $p = \left[\sqrt{100}\right] = 10$ and $p = \left[\sqrt{1000}\right] = 31$, respectively, does not seem entirely implausible. Furthermore, for the cosine window, the average cut-off values in these cases are just under 10 and just over 35, respectively (see Table 1.b). What is somewhat surprising, however, is that the nonparametric estimators seem to outperform the parametric ones.

The tuning parameters controlling the smoothness of the estimate have a different interpretation in the nonparametric case than in the parametric case. The smaller the order $p$ of the AR model, the smoother the result, whereas in periodogram smoothing, a larger value of the cut-off value $r$ yields a smoother result than a smaller one (see Table 1.b).

Table 1: Results of simulation study

(a) Average of goodness-of-fit statistics (24) over 1,000 realizations of length $n = 200$ or $n = 2000$ obtained from MA processes (22) with $\theta = -0.5$, 0.5 and $q = 1,\ldots,6$.

| | | $\theta = -0.5$ | | | | | | $\theta = 0.5$ | | | | | |
|---|---|---|---|---|---|---|---|---|---|---|---|---|---|
| $n$ | q | rect | tri | cos | $AR$ | $AR_{fix}$ | $cos_{fix}$ | rect | tri | cos | $AR$ | $AR_{fix}$ | $cos_{fix}$ |
| 200 | 1 | .035 | .039 | .039 | .043 | .056 | .046 | .033 | .037 | .036 | .041 | .054 | .044 |
| | 2 | .039 | .042 | .040 | .051 | .056 | .045 | .043 | .045 | .043 | .051 | .057 | .047 |
| | 3 | .046 | .047 | .046 | .058 | .058 | .046 | .045 | .046 | .045 | .056 | .057 | .046 |
| | 4 | .049 | .050 | .049 | .062 | .060 | .047 | .051 | .051 | .051 | .063 | .060 | .049 |
| | 5 | .054 | .054 | .053 | .067 | .061 | .050 | .054 | .053 | .053 | .066 | .060 | .050 |
| | 6 | .056 | .055 | .055 | .069 | .070 | .052 | .058 | .057 | .056 | .069 | .070 | .054 |
| 2000 | 1 | .013 | .019 | .015 | .015 | .029 | .025 | .013 | .020 | .015 | .014 | .029 | .025 |
| | 2 | .016 | .022 | .018 | .018 | .029 | .025 | .016 | .023 | .018 | .019 | .029 | .025 |
| | 3 | .019 | .024 | .020 | .022 | .029 | .025 | .018 | .023 | .020 | .021 | .029 | .025 |
| | 4 | .020 | .025 | .021 | .023 | .029 | .025 | .020 | .025 | .022 | .024 | .029 | .025 |
| | 5 | .022 | .026 | .023 | .026 | .030 | .025 | .022 | .026 | .023 | .026 | .030 | .025 |
| | 6 | .023 | .027 | .024 | .027 | .029 | .025 | .024 | .027 | .025 | .027 | .030 | .025 |

(b) Average value of the smoothing parameter (AR order p in the parametric case and cut-off value r in the nonparametric case) over 1,000 realizations.

| | | $\theta = -0.5$ | | | | | | $\theta = 0.5$ | | | | | |
|---|---|---|---|---|---|---|---|---|---|---|---|---|---|
| $n$ | q | rect | tri | cos | $AR$ | $AR_{fix}$ | $cos_{fix}$ | rect | tri | cos | $AR$ | $AR_{fix}$ | $cos_{fix}$ |
| 200 | 1 | 14.5 | 15.5 | 17.1 | 3.4 | 10 | 10 | 14.9 | 15.7 | 17.6 | 3.2 | 10 | 10 |
| | 2 | 11.2 | 12.5 | 15.1 | 5.3 | 10 | 10 | 11.2 | 12.8 | 15.4 | 5.1 | 10 | 10 |
| | 3 | 8.6 | 10.1 | 12.3 | 6.7 | 10 | 10 | 8.5 | 10.1 | 12.2 | 6.7 | 10 | 10 |
| | 4 | 7.2 | 8.7 | 10.4 | 7.7 | 10 | 10 | 7.2 | 8.8 | 10.6 | 7.8 | 10 | 10 |
| | 5 | 6.3 | 7.8 | 9.2 | 8.9 | 10 | 10 | 6.1 | 7.7 | 9.1 | 8.8 | 10 | 10 |
| | 6 | 5.6 | 7.1 | 8.3 | 9.6 | 10 | 10 | 5.7 | 7.2 | 8.4 | 9.8 | 10 | 10 |
| 2000 | 1 | 99.9 | 80.8 | 128.5 | 5.2 | 31 | 31 | 101.6 | 76.6 | 128.9 | 5.0 | 31 | 31 |
| | 2 | 62.1 | 50.2 | 79.4 | 8.5 | 31 | 31 | 60.8 | 49.2 | 78.4 | 8.8 | 31 | 31 |
| | 3 | 46.2 | 40.2 | 60.5 | 12.1 | 31 | 31 | 45.7 | 39.8 | 60.0 | 11.9 | 31 | 31 |
| | 4 | 37.0 | 33.5 | 48.4 | 14.8 | 31 | 31 | 37.4 | 33.0 | 48.9 | 15.1 | 31 | 31 |
| | 5 | 31.8 | 29.1 | 41.5 | 18.0 | 31 | 31 | 31.5 | 28.9 | 41.4 | 17.8 | 31 | 31 |
| | 6 | 27.7 | 26.2 | 36.6 | 20.5 | 31 | 31 | 27.5 | 26.1 | 36.5 | 20.6 | 31 | 31 |

## 5. Discussion

The misconception that nonparametric estimators are generally much more volatile than parametric ones stems, on the one hand, from the fact that the degree of their smoothness is widely determined subjectively, and people in general - not just stock analysts - tend to see patterns where none exist, and on the other hand, on the fact that in cases where one believes in a "true" model, consistent model selection methods such as BIC are often used, and these typically select much smaller model dimensions than, for example, AIC. However, the results of our study show that in situations where one prefers to use AIC over BIC because overfitting is safer there than underfitting, the nonparametric estimates do not fundamentally differ from the parametric ones when a criterion related to AIC is used for them as well. The availability of alternatives opens up the possibility of selecting the “best” estimate (albeit, admittedly, subjectively) from a range of reasonable estimates - all of which were obtained fully automatically - rather than blindly relying on a single automatic estimation method. The practical benefit should become clear when the attempt to select the best estimate in each of the applications in Section 3 (see Figures 3–5) does not invariably lead to the choice of the AR estimate. Choosing from a very small number of estimates is often less difficult and less arbitrary than one might assume. On the one hand, some estimates might hardly differ from one another, making it completely irrelevant which one we choose; on the other hand, some estimates might stand out in a negative way, allowing them to be eliminated from the outset - such as the volatile rectangular-window estimate in Figure 5.d or the ARMA spectral densities with extreme spikes. Of course, not every spike is necessarily a bad thing; it can also provide valuable information, such as an indication of an overlooked seasonality or a botched seasonal adjustment.

In the end, it is not even always necessary to settle for just one option. Sometimes, simply gaining an overview of the range of plausible options can be enough - for example, when the spectral analysis is just a small part of a much larger project and the goal is to explore various scenarios. When designing a climate model, for instance, a cross-spectral analysis could help determine the specific lag with which an exogenous variable should be included in the model. For that, we would only need a specific section of the phase spectrum, hence two different estimates might result in the same lag. On the other hand, if one estimate implies a lag of 2 and another implies a lag of 3, then this would likely have to be taken into account when assessing the uncertainty of a forecast obtained with this climate model.

## References

Akaike, H. (1969) Fitting autoregressive models for prediction. *Annals of the Institute of Statistical Mathematics*, 21, 243-247. https://doi.org/10.1007/BF02532251

Akaike, H. (1973) Information theory and an extension of the maximum likelihood principle. In: Petrov, B.N., Csaki, F. (eds.), *Second International Symposium on Information Theory*, Akademia Kiado, Budapest, 267-281.

Beltrao, K.I., Bloomfield, P. (1987) Determining the bandwidth of a kernel spectrum estimator. *Journal of Time Series Analysis*, 8,1, 21-38 https://doi.org/10.1111/j.1467-9892.1987.tb00418.x

Geweke, J., Porter-Hudak, S. (1983) The estimation and application of long memory time series models. *Journal of Time Series Analysis*, 4, 221–38. http://dx.doi.org/10.1111/j.1467-9892.1983.tb00371.x

Hurvich, C.M. (1985) Data-driven choice of a spectrum estimate: extending the applicability of cross-validation methods. *Journal of the American Statistical Association*, 80, 392, 933–940. https://doi.org/10.1080/01621459.1985.10478207

Kabaila, P. (2002). On Variable Selection in Linear Regression. *Econometric Theory*, 18, 4, 913–925. https://doi.org/10.1017/S0266466602184052

Lee, T.C.M. (1997) A simple span selector for periodogram smoothing, *Biometrika*, 84, 4, 965–969. https://doi.org/10.1093/biomet/84.4.965

Morice, C.P., Kennedy, J.J., Rayner, N.A., Winn, J.P., Hogan, E., Killick, R.E., Dunn, R.J.H., Osborn, T.J., Jones, P.D., Simpson, I.R. (2021) An updated assessment of near-surface temperature change from 1850: the HadCRUT5 dataset. *Journal of Geophysical Research*, 126, e2019JD032361. https://doi.org/10.1029/2019JD032361

R Core Team (2024) R: A Language and Environment for Statistical Computing. R Foundation for Statistical Computing, Vienna, Austria. https://www.R-project.org/

Reschenhofer, E. (1999) Improved estimation of the expected Kullback-Leibler discrepancy in case of misspecification. *Econometric Theory*, 15, 3, 377-387. https://doi.org/10.1017/S0266466699153052

Reschenhofer, E. (2015) Consistent variable selection in large regression models. *Journal of Statistics: Advances in Theory and Applications*, 14, 1, 49-67. http://dx.doi.org/10.18642/jsata_7100121536

Reschenhofer, E. (2026) Remarks on the acceleration of global warming and the imminent breach of the 1.5°C Paris Agreement target. *arXiv*: 2604.10694. https://doi.org/10.48550/arXiv.2604.10694

Robinson, P. M. (1991). Automatic frequency domain inference on semiparametric and nonparametric models. *Econometrica*, 59, 5, 1329-1363. https://www.jstor.org/stable/2938370

Rothman, D. (1968) Letter to the editor. *Technometrics*, 10, 432. https://doi.org/10.2307/1267069

Sawa, T. (1978) Information criteria for discriminating among alternative regression models. *Econometrica*, 46, 6, 1273-1291. https://doi.org/10.2307/1913828

Schwarz, G. (1978) Estimating the dimension of a model. *The Annals of Statistics*, 6, 461-464. https://www.jstor.org/stable/2958889

Shibata, R. (1980) Asymptotically efficient selection of the order of the model for estimating parameters of a linear process. *Annals of Statistics*, 8, 147-164. https://www.jstor.org/stable/2240749

Shibata, R. (1981) Optimal selection of regression variables. *Biometrika*, 68, 1, 45-54. https://doi.org/10.2307/2335804

Sugiura, N. (1978) Further analysis of the data by Akaike's information criterion and the finite corrections. *Communications in Statistics, Theory and Methods*, A7, 13-26. https://doi.org/10.1080/03610927808827599

Wahba, G. (1980) Automatic Smoothing of the Log Periodogram. *Journal of the American Statistical Association*, 75, 369, 122–132. https://doi.org/10.1080/01621459.1980.10477441